\documentstyle[12pt]{article}

\newcommand{\bce}{\begin{center}}
\newcommand{\ece}{\end{center}}
\newcommand{\beq}{\begin{equation}}
\newcommand{\eeq}{\end{equation}}
\newcommand{\bea}{\vspace{0.25cm}\begin{eqnarray}}
\newcommand{\eea}{\end{eqnarray}}

\newcommand{\ba}{\begin{array}}
\newcommand{\ea}{\end{array}}

\newcommand{\doublespace}{
    \renewcommand{\baselinestretch}{1.6}\large\normalsize}

\def\lsim{\mathrel{\rlap{\lower4pt\hbox{\hskip1pt$\sim$}}
    \raise1pt\hbox{$<$}}}         
\def\gsim{\mathrel{\rlap{\lower4pt\hbox{\hskip1pt$\sim$}}
    \raise1pt\hbox{$>$}}}         

\def\Pom{{\bf I\!P}}

\def\beq{\begin{equation}}
\def\endeq{\end{equation}}
\def\arr{\begin{eqnarray}}
\def\endarr{\end{eqnarray}}
\makeindex

  \advance\oddsidemargin  by -1in
  \advance\evensidemargin by -1in
  \advance\topmargin      by -0.5in
\begin{document}

\doublespace

\begin{center}
{\LARGE \bf
On the determination of double diffraction dissociation cross section
at HERA
\bigskip\\ }
{H.Holtmann$^{1}$, N.N.Nikolaev$^{1,2}$, A.Szczurek$^{3}$,
J.Speth$^{1}$ and B.G.Zakharov$^{2}$}
\medskip\\
{\sl
$^{1}$IKP(Theorie), Forschungszentrum  J\"ulich GmbH.,\\
52425 J\"ulich, Germany \\
$^{2}$  Landau Institute for Theoretical Physics, \\
 GSP-1, 117940, ul.Kosygina 2, 117334 Moscow, Russia \\
$^{3}$ Institute of Nuclear Physics,\\
ul. Radzikowskiego 152, PL-31-342 Krak\'ow, Poland}

\vskip 1cm

{\large \bf Abstract}

\end{center}

The excitation of the proton into undetected multiparticle states
(double diffraction dissociation) is an important background
to single diffractive deep-inelastic processes
$ep \rightarrow e'p'\rho^{0}, ~e'p'J/\Psi, ~e'p'X$
at HERA.
We present estimates of the admixture of the double diffraction
dissociation events in all diffractive events. We find that
in the $J/\Psi$ photoproduction, electroproduction of the $\rho^0$
at large $Q^2$ and diffraction dissociation
of real and virtual photons into high mass states $X$
the contamination of the double diffraction dissociation
can be as large as $\sim 30\%$,
thus affecting substantially the experimental tests of the pomeron
exchange in deep inelastic scattering at HERA.
We discuss a possibility of tagging the
double diffraction dissociation by neutrons observed in the
forward neutron calorimeter.
We present evaluations of the spectra of neutrons and
efficiency of neutron tagging based on the experimental data
for diffractive processes in the proton-proton collisions.

\pagebreak

\section {Introduction}

Important information on the pomeron exchange in deep inelastic
scattering (DIS) comes from diffractive (large rapidity gap)
DIS of Fig.1a,
\beq
\gamma^{*}+p\rightarrow X+p' \, ,
\label{eq:1}
\endeq
at small $x={Q^{2}\over (Q^{2}+W^{2})} \ll 1$, which is being
actively studied at HERA \cite{D93,H1}. $X$ here can be either
vector mesons $\rho^{0},\phi,J/\Psi,\Upsilon, ...,$ or
the low-mass (LM), $M_{X}^{2}\sim  Q^{2}$,  and high-mass (HM),
$M_{X}^{2} \gg Q^{2}$, continuum excitations of the photon.
In the single diffraction (SD) reaction (\ref{eq:1}),
the final-state proton $p'$ is separated from the hadronic
debris $X$ of the photon by a large (pseudo)rapidity gap
$ \Delta \eta \approx -\ln{x_{\Pom}}$,
where $x_{\Pom}={(M^{2}_{X}+Q^{2})/(W^{2}+Q^{2})}\ll 1$ is
the fraction of the proton's momentum taken away by the pomeron.
At the present stage of the H1 and ZEUS detectors,
the recoil proton $p'$ with very small energy loss cannot be registered
and the SD events of Fig.~1a could not be experimentally
distinguished from the double diffraction dissociation (DD) events
of Fig.~1b, $ \gamma^{*}+p\rightarrow X+Y $, in which proton
resonances and/or multiparticle states $Y$
escape undetected into the beam pipe.
The experimental separation of DD at HERA is
very important for several reasons. A clean signal of single diffraction
of real and virtual photons is important for
understanding the $Q^{2}$-evolution of the pomeron exchange in
diffractive reactions and for testing models of the pomeron
structure function
\cite{IS85,FS85,BCSS87,DL87,Nikolaev,LW94,MP94,C94,CKMT95,BW94}.
A comparison of double and single diffraction
will provide an unique information on the $Q^{2}$-dependence of
the factorization properties of the pomeron and sheds light on the
much disputed issue of whether the soft pomeron exchange at
small $Q^{2}$ will be superceded by hard pomeron exchange
at larger $Q^{2}$. There are interesting predictions
for the cross section and the diffraction slope of pomeron
exchange ractions
$\gamma (\gamma^{*}) + p \rightarrow \rho^{0} (J/\Psi) + p$,
at HERA \cite{DL87b,C90,R93,NNZ94,B94,NZZ94,NZZ95}.
The tests of these pedictions require a separation
of the background from DD reaction
$\gamma (\gamma^{*}) + p \rightarrow \rho^{0} (J/\Psi) + Y$,
which can significantly affect the diffraction slope measurements.

The purpose of this paper is twofold. Firstly, following
microscopic models, in which the properties of the pomeron do not
change significantly from diffraction of hadrons and real photons
to diffractive DIS (see for instance: \cite{DL87,Nikolaev,CKMT95}),
we present estimates of the DD background to large rapidity gap
and exclusive vector meson production in DIS and real photoproduction,
based on the factorization approximation and huge body of
experimental information on diffractive hadron-hadron interactions.
We find that in the $J/\Psi$ production
and diffraction dissociation of photons into high mass states
$X$, the ratio of the DD background to the SD production is very large,
$\gsim 30\% $.
This is of particular importance in the interpretation of the
$t$-dependence of the $J/\Psi$ production (or the $\rho^0$, $\omega$,
$\phi$ production at large $Q^2$).
Secondly, we explore in some detail the possibility of
tagging DD events by neutrons contained in the proton
excitations $Y$. These neutrons can easily be detected
in the forward neutron calorimeter (FNC) \cite{FNC}. We find
that an approximately $\kappa_{n} \sim 25\%$ fraction of
double diffraction dissociation events contains a neutron which
satisfies a typical neutron energy cut of $z=E_{n}/E_{p} > z_{min} \sim
0.5$. We conclude that the neutron tagging of DD is
feasible, and the observed fraction $f_{DD}^{(n)}$  of
neutron tagged diffractive DIS will
allow a direct experimental determination of the overall
fraction $f_{DD}$ of diffractive DIS which proceeds via double
diffraction dissociation:
$f_{DD} = {f_{DD}^{(n)}/\kappa_{n}}.$
The experimental implementation of the neutron tagging of
DD shall shed a light on many facets of double diffraction dissociation
and the pomeron exchange in DIS.

\vskip 1cm

\section{Regge factorization and single and
 double diffraction dissociation}

Our analysis of DD is based on the well established similarity
of the mass spectrum of hadronic states $Y$ in the double diffraction
dissociation process $p+p\rightarrow X+Y$ of Fig.~1d and
in the single diffraction dissociation process
$p+p\rightarrow p+Y$ of Fig.~1c, in conformity with the
Regge theory factorization for single pomeron exchange:
\begin{equation}
\left.{d\sigma(pp\rightarrow XY)\over dM_{X}^{2}dM_{Y}^{2}
 dt} \right|_{t=0}
=G(p\rightarrow X)\cdot G(p\rightarrow Y) \; .
\label{eq:2}
\end{equation}
In proton-proton interactions, this factorization law
has been shown to hold to an accuracy $\lsim$ 20\%
(for a detailed discussion see \cite{K79,AG81,G83,ZZ88}).
Evaluation of the admixture of DD to large rapidity gap events,
$$
f_{DD}^{ap}={\sigma_{DD}(ap\rightarrow XY) \over
\sigma_{SD}(ap\rightarrow Xp')+
\sigma_{DD}(ap\rightarrow XY)}
$$
using the factorization (\ref{eq:2}) involves integration of the DD
cross section over $t$ and $M_{Y}^2$ and requires a knowledge of
the diffraction slope $b_{DD}(XY)$.
For reactions with hadrons and real photons
($a=p,\pi,K,\gamma$; the target is always assumed to be a proton)
the gross features of $b_{ap \rightarrow DD}(XY)$ are consistent
with the Regge theory relationship
\beq
b_{DD}(ap\rightarrow XY)\approx
b_{SD}(ap\rightarrow Xb)+b_{SD}(ap \rightarrow aY)-
b_{el}(ap)
\label{eq:3}
\endeq
and can be described as follows (see also \cite{K79,AG81,G83,ZZ88}).
In the single diffraction excitation of the beam hadron $a$
or the target proton or in double diffraction dissociation of
both the beam and target hadrons into
low-mass (LM) states $X_{LM}$ and $Y_{LM}$
\beq
b_{DD}(ap\rightarrow X_{LM}+Y_{LM})\sim
b_{SD}(ap\rightarrow X_{LM}+p)\sim
b_{SD}(ap \rightarrow a+Y_{LM})\sim
b_{el}(ap) \; .
\label{eq:4}
\endeq
In single diffraction dissociation
when the beam hadron (or target proton) is excited
into the high-mass (HM) continuum states, the diffraction
slope, as a function of the mass $M_{X}$ and $M_{Y}$,
levels off at a value $B_{SD}$:
\beq
b_{SD}(ap\rightarrow X_{HM}+p)
\sim b_{SD}(ap \rightarrow a+Y_{HM}) \equiv B_{SD}
\sim {1\over 2}
b_{el}(pp).
\label{eq:5}
\endeq
In double diffraction dissociation when both the beam hadron
and the target proton are excited into HM states $X$ and $Y$,
the diffraction slope is also approximately constant, but considerably
smaller than $B_{SD}$:
\beq
b_{DD}(pp\rightarrow X_{HM}+Y_{HM}) \equiv B_{DD}
\approx 2 \,GeV^{-2}\, ,
\label{eq:6}
\endeq
in agreement with (\ref{eq:3})-(\ref{eq:5}).
The well known consequence of the factorization (\ref{eq:2})
for the fixed masses X and Y in the exponential approximation
of the $t$-dependence is \cite{K79,AG81,G83}
\begin{equation}
\sigma_{DD}(ap \rightarrow XY) \approx
{\sigma_{SD}(ap \rightarrow aY)\cdot
\sigma_{SD}(ap \rightarrow Xp) \over
\sigma_{el}(ap)}\cdot
{b_{SD}(ap \rightarrow aY)\cdot b_{SD}(ap \rightarrow Xp) \over
b_{DD}(ap \rightarrow XY)\cdot
b_{el}(ap)} \; .
\label{eq:7}
\end{equation}
For an approximate estimation of the SD and DD cross sections,
we integrate over the excitations $X$ and $Y$ splitting
the mass spectra
into the low-mass and high-mass regions and making use of
(\ref{eq:3})-(\ref{eq:7}).
In hadronic reactions the boundary
between the low-mass (resonances, ...) and high-mass (the
continuum) regions is $M_{X,Y}\sim 2\,GeV$ \cite{K79,AG81,G83,ZZ88}.

In diffractive DIS and in photoproduction of heavy quarkonia
the situation is somewhat more complicated.
Let us consider first the case of real photoproduction.
Following the established tradition, we will refer
to the single diffractive photoproduction
$\gamma p \rightarrow V p$ ($V = \rho^{0}, \omega, \Phi, J/\Psi$)
as "elastic" production and to the double diffraction
$\gamma p \rightarrow V Y$ as "inelastic" production.
It is known experimentally \cite{A92} that for elastic production
$b_{el}(\gamma p \rightarrow \rho^0 p) \approx b_{el}(\pi p)$
(for $\omega$ production see also \cite{BCD81}).
It is natural to expect that:
\begin{eqnarray}
b_{in}(\gamma p \rightarrow \rho^0 Y_{LM}) \approx
b_{el}(\gamma p \rightarrow \rho^0 p) \approx b_{el}(\pi N) \; , \\
b_{in}(\gamma p \rightarrow \rho^0 Y_{HM}) \approx B_{SD} \; .
\end{eqnarray}
In elastic production of heavy quarkonia
$\gamma p \rightarrow J/\Psi\,p$,
as well as for deeply virtual (large $Q^2 \gsim m_{J/\Psi}^2$)
photoproduction of light vector mesons, the contribution from
the vertices
$\gamma \rightarrow J/\Psi$ and
$\gamma \rightarrow \rho^0$
to the diffraction slope becomes small \cite{NZZ95}
\begin{equation}
\Delta b(Q^2) \approx {9 \over 2(m_{V}^2 + Q^{2})} \ll b_{el}(pp) \; .
\label{Deltab}
\end{equation}
For these reasons we expect:
\begin{eqnarray}
b_{el}(\gamma p \rightarrow J/\Psi \, p) \approx
b_{in}(\gamma p \rightarrow J/\Psi \, Y_{LM}) \approx B_{SD} \; , \\
b_{in}(\gamma p \rightarrow J/\Psi \, Y_{HM}) \approx B_{DD} \; .
\end{eqnarray}
For the photoproduction of $\rho^0$
at intermediate $Q^2 \gsim 2 GeV^2$ we expect a smooth interpolation
\begin{eqnarray}
b_{in}(\gamma p \rightarrow \rho^{0} Y_{LM})
\sim B_{SD} + \Delta b(Q^2) \; , \\
b_{in}(\gamma p \rightarrow \rho^{0} Y_{HM})
\sim B_{DD} + \Delta b(Q^2) \; .
\end{eqnarray}

Let us consider now diffractive DIS.
Let $\beta$ be the Bjorken variable for DIS on the pomeron:
$\beta = x/x_{\Pom} = Q^{2}/(Q^{2}+M_{X}^{2})$.
It has been argued \cite{Nikolaev} that the LM region corresponds
to the interaction of $\gamma^{*}$ with the quark-antiquark
valence component of the pomeron
($\beta \gsim 0.2$, i.e. $M_{X}^{2} \lsim 5 Q^{2}$).
In this case the educated guess is
\begin{eqnarray}
b_{SD}(\gamma^{*} p \rightarrow X_{LM} \, p) \approx
b_{DD}(\gamma^{*} p \rightarrow X_{LM} Y_{LM}) \approx
b_{el}(\pi N) \; , \\
b_{DD}(\gamma^{*} p \rightarrow X_{LM} Y_{HM}) \approx B_{SD} \; .
\end{eqnarray}
The HM region corresponds to the values of $\beta \lsim 0.2$,
when $\gamma^{*}$ probes the sea of the pomeron generated
by the valence gluon-gluon component of the pomeron \cite{Nikolaev}.
Here one expects the diffractive slopes
\begin{eqnarray}
b_{SD}(\gamma^* p \rightarrow X_{HM} \, p) \approx
b_{DD}(\gamma^* p \rightarrow X_{HM} Y_{LM}) \approx B_{SD} \; , \\
b_{DD}(\gamma^* p \rightarrow X_{HM} Y_{HM}) \approx B_{DD} \; .
\end{eqnarray}
The prescriptions (8-18) for the diffraction slope in SD and DD,
for real and virtual photoproduction are educated guesses based
on the assumption that the pomeron properties do not depend strongly
on flavour and/or $Q^2$ and should be a subject of experimental tests.
First of all, even in the best studied proton-proton interactions
the value of $B_{DD}$ (Eq.(\ref{eq:6}))  comes from studying only some
exclusive channels. Secondly, it is not excluded that $B_{DD}$
in DIS is slightly different from what is known from hadronic reactions.

\vskip 1cm

\section {Double diffraction dissociation at HERA}

In the case of the ZEUS or H1 collaboration main calorimeters,
the hole for the beam pipe determines the minimal polar scattering
angle $\theta_m$, and the corresponding maximal pseudorapidity
$\eta_{F} = log {2 \over \theta_m}$,
of the detected secondary particles.
The range of masses $M_{Y}$ of the undetected excitations $Y$
of the proton can be evaluated as follows:
If the proton excites into a state of mass $M_{Y} \gg m_{p}$,
then the held back pions with the transverse
momentum $k_{\perp}$, transverse mass
$\mu_{\perp}=\sqrt{\mu_{\pi}^{2} + k_{\perp}^{2}}$,
longitudinal momentum $k_{z}\approx -{1\over 2}M_{Y}$ and
rapidity in the co-moving frame
$
y  \approx -\log(2|k_{z}|/\mu_{\perp}) \approx
- - -\log(M_{Y}/\mu_{\perp})
$,
have the smallest (pseudo)rapidity $\eta$ in the HERA reference frame
\beq
\eta=\log{2E_{p}\over M_{Y}}+y
\approx \log{2E_{p}\mu_{\perp}^{2} \over M_{Y}^{2} } \approx
\log{2E_{p}k_{\perp}^{2} \over M^{2} }  \, .
\label{eq:8}
\endeq
The forward calorimeter of the ZEUS collaboration misses secondary
particles with $\eta \geq \eta_{F}=4.3$ \cite{D93}.
Then, for the proton energy $E_{p}=820\,GeV$ and typical
$k_{\perp} \sim 0.5\,GeV/c$ all the secondary hadrons coming
from excitations of the proton with
\begin{equation}
M_{Y} <
M_{max} \approx \sqrt{2E_{p}k_{\perp}}\exp(-{1\over 2}\eta_{F})
\sim 4\, GeV
\label{eq:9}
\end{equation}
will escape into the beam pipe. Monte Carlo studies for the
ZEUS detector suggest $M_{max} \approx $5\,$GeV$ \cite{Doecker}, which
we use in the further analysis. The H1 detector has a similar
angular acceptance \cite{H1}.

In the following we will use the factorization relation (\ref{eq:7})
to estimate the double diffraction dissociation background
to single diffraction dissociation in DIS (Fig. 1a) in terms of
SD cross sections in proton-proton collisions. We start with a brief
reminder of salient features of the single diffraction in
$p+p\rightarrow  p'+Y$.
The key quantity here is $d\sigma_{SD} \over dt dM_{Y}^{2}$,
integration of which gives:
\begin{equation}
\sigma_{SD}(M_{Y}\leq M_{max})=
\int^{M_{max}^{2}} dM_{Y}^{2} \int dt
{d\sigma_{SD} \over dt dM_{Y}^{2}} \; .
\label{eq:10}
\end{equation}
Although the diffraction slope $b_{SD}(M_{Y}^{2})$ exhibits a nontrivial
mass dependence,
the quantity
\begin{equation}
\Sigma_{SD}=
 M_{Y}^{2} \int dt \,
 {d^{2}\sigma_{SD} \over dt dM_{Y}^{2}}
\end{equation}
is practically flat, $\Sigma_{SD} = 0.68 \pm 0.05 \,mb$,
in a broad range of excited masses, $1.7 \lsim M_{Y}^{2} \lsim
30\,GeV^{2}$, and decreases towards smaller $M_{Y}$ \cite{A76,G83}.
Compared to Eq.(45) in \cite{G83} we neglect the finite-energy
corrections. This leads to the simple paprametrization
\begin{equation}
\sigma_{SD}(pp \rightarrow pY; M_{Y}^{2} < M_{max}^{2}) \approx
2\Sigma_{SD}\log\left({M_{max} \over 1.25(GeV)}\right) \, ,
\label{eq:11}
\end{equation}
which gives 1.9 $mb$ (2.1 $mb$) for $M_{max} = 5\,GeV$
($M_{max} = 6\,GeV$).
The corresponding LM/HM partition is
$\sigma_{SD}(LM) \approx 0.64\,mb$ and
$\sigma_{SD}(HM) \approx 1.25\,mb$ ($1.5\,mb$), with an uncertainty of
$\lsim 20\%$. Consequently, for the LM excitations of the beam proton
$X_{LM}$, summing over all the excitations of the target proton $Y$,
we obtain by virtue of Eq.(\ref{eq:7})
\beq
f_{DD}^{pp}(X_{LM}) \equiv
{ \sigma(pp \rightarrow X_{LM} Y )
\over
\sigma(pp \rightarrow X_{LM},p) + \sigma(pp \rightarrow X_{LM},Y) }
\approx
{\sigma_{SD}(pp \rightarrow pY) \over
\sigma_{el}(pp) +
\sigma_{SD}(pp \rightarrow pY)} \approx 0.21\, .
\label{fDDLM}
\endeq
For the HM excitations of the beam proton $X_{HM}$, care must
be taken of the difference between $B_{SD}$ and $B_{DD}$
when integrating over masses of the target proton debris $Y$:
\beq
f_{DD}^{pp}(X_{HM}) \equiv
{ \sigma(pp \rightarrow X_{HM} Y )
\over
\sigma(pp \rightarrow X_{HM},p) + \sigma(pp \rightarrow X_{HM},Y) } \\
\approx \\
{\sigma_{SD}(LM)+
 \chi_{DD}\sigma_{SD}(HM)
 \over
\sigma_{el}(pp) + \sigma_{SD}(LM) + \chi_{DD}\sigma_{SD}(HM) }
\approx 0.26\, ,
\label{fDDHM}
\endeq
where $\chi_{DD}$ here is a new parameter defined as
\begin{equation}
\chi_{DD} = {B_{SD} \over B_{DD}} \cdot {B_{SD} \over b_{el}} \, .
\end{equation}
In the evaluation above we have used $\sigma_{el}(pp) \approx 7.3\,mb$
and $\chi_{DD}\approx 1.5$. If $M_{max} = 6\,GeV$, we find
a somewhat larger
$f_{DD}^{pp}(X_{LM})\approx 0.23$ and $f_{DD}^{pp}(X_{HM})\approx 0.28$.
The very small value of the slope in DD processes
(see Eq. (\ref{eq:6}) ) leads to an enhanced effect of excitation of
the target proton in high mass states $Y_{HM}$ and to
a large admixture $f_{DD}^{pp}(X_{HM})$. Please note that
$f_{DD}(X_{HM})$ depends crucially on the poorly known $B_{DD}$,
of which direct measurement is yet lacking, and still larger
values of $\chi_{DD}$ and as a consequence larger values of
$f_{DD}(X_{HM})$ are possible.

The structure function of the pomeron is operationally
defined through the cross section for single diffraction dissociation
of virtual photons \cite{IS85,FS85,BCSS87}.
In the present HERA experiments
the cross section for double diffraction dissociation constitutes
an unwanted background.
Therefore the extraction of the SD cross section from large
rapidity gap data requires a correction for the DD background.
Extrapolation of the above considerations for hadronic reactions
to diffractive DIS requires the assumption that
properties of the pomeron exchange do not change from
hadronic scattering to diffractive DIS.
Under this assumption we find that
$f_{DD}(\gamma^{*} \rightarrow X_{LM})$ and
$f_{DD}(\gamma^{*} \rightarrow X_{HM})$
will be approximately independent
of $x_{\Pom}$ and $Q^{2}$:
\begin{eqnarray}
f_{DD}(\gamma^* \rightarrow X_{LM}) \approx
f_{in}(\gamma \rightarrow \rho^{0}) \approx
f_{DD}^{pp}(X_{LM}) \approx 0.21-0.23 \; , \\
f_{DD}(\gamma^* \rightarrow X_{HM}) \approx
f_{in}(\gamma \rightarrow J/\Psi) \approx
f_{DD}(\gamma^* \rightarrow \rho^{0}, Q^2 \gsim M_{J/\Psi}^2) \approx
f_{DD}^{pp}(X_{HM}) \approx 0.26-0.28 \; .
\end{eqnarray}

Notice that the contribution from the DD background in
the HM (small-$\beta$) region is larger than in
the LM (large-$\beta$) region. A correct interpretation of the
experimentally observed $\beta$ distributions in terms
of the pomeron structure function will require a careful correction
for the $\beta$-dependence of the background.
Experimental determination of the
$(x_{\Pom},Q^{2})$-dependence of $f_{DD}$ would be an
extremely interesting test of the current ideas on the pomeron.

Similarly, in the photo- and electroproduction of the
$J/\Psi$ (and heavier quarkonia) the expected "inelastic"
background
$f_{in}(\gamma \rightarrow J/\Psi) \sim f_{DD}^{pp}(X_{HM})
\approx 0.26-0.28$ is larger than in real photoproduction
of $\rho^{0}$, which is consistent with the H1 data \cite{H1}.
Such a large admixture of inelastic events with a very small
diffraction slope $B_{in}(J/\Psi) \sim B_{DD}$ can substantially
affect measurements of the energy dependence of the slope
$B_{el}(J/\Psi)$, which is very important for testing
the BFKL pomeron.
In the photoproduction of $\rho^{0}$, the background
from "inelastic" production is weaker and the slopes of
the "elastic" and "inelastic" components differ less:
$B_{el}(\rho^{0}) \approx b_{el}(\pi N)$ \cite{A92} vs. the expected
$B_{in}(\rho^{0}) \approx B_{SD}$. According to Eq.(\ref{Deltab})
the decrease of $\Delta b(Q^2)$ with $Q^2$ leads to an increase of
the admixture of inelastic $\rho^{0}$ production with $Q^2$,
which can be described by the $Q^2$-dependence of the parameter
$\chi_{DD}$ in the counterpart of Eq.(\ref{fDDHM})
\begin{equation}
\chi_{DD}(Q^2) \approx
{[B_{SD} + \Delta b(Q^2)]^2 \over [B_{DD} + \Delta b(Q^2)]\,b_{el}} \; .
\end{equation}
We expect that the $f_{in}(\gamma^* \rightarrow \rho^0)$
rises with $Q^2$ from $f_{in}(\gamma \rightarrow \rho^0)$ for real
photoproduction up to 0.28 at large $Q^2$. This prediction could be
easily tested at HERA.

\vskip 1cm

\section {Neutron tagging of double diffraction dissociation at HERA}

The strong admixture (Eqn.(26,27)) of DD processes in diffractive
DIS makes its direct determination very important issue. A direct
detection of charged particles from double diffraction dissociation
is hardly possible. On the other hand, observation of very forward
neutrons is easier \cite{FNC}.
Therefore in the following we wish to discuss
the feasibility of neutron tagging in more detail.

The interpretation of the data on neutron tagged DD requires
knowledge of the neutron content for specific final states $Y$.
It is known from hadronic reactions that to a good approximation
the spectrum of the final state $Y$ can be approximated
in terms of three distinct components.
The first prominent component is $Y=\pi N$
with the cross section
$\sigma_{1}= \sigma_{SD}(pp \rightarrow p+(\pi N)) \approx (0.4-0.45)\,
mb$.
This is the predominantly LM channel, because the observed
$M_{Y}$ spectrum has a characteristic broad bump which peaks
at $M_{Y}\sim 1.5\,GeV$, extends up to $M_{Y} \lsim 1.8\,GeV$ and
is followed by a tail up to $M_{Y} \lsim 2.4\,GeV$ \cite{B78,AG81}.
Since the diffractively excited $\pi N$ system is in isospin
${1\over 2}$ state, the average multiplicity of neutrons in
the $\pi N$ final state $\langle n_{n}^{(1)} \rangle= {2\over 3}$.
The second prominent, predominantly LM channel, is $Y=N\pi\pi$.
It is dominated by $\Delta \pi$ production \cite{W75,C80}.
The measured cross section is
$\sigma_{SD}(pp \rightarrow p+(p\pi^{+}\pi^{-}))
\approx $(0.15-0.2)$\,mb$.
The observed mass spectrum has a broad peak at $M_{Y}\sim 1.7\,GeV$
and a tail up to $M_{Y}\lsim 2.5\,GeV$ \cite{K79,AG81}.
A simple isospin analysis leads to
\arr
\sigma_{2}=
\sigma_{SD}(pp \rightarrow p+(N\pi\pi))
\approx {9\over 5}
\sigma_{SD}(pp \rightarrow p+(p\pi^{+}\pi^{-})) =
(0.27-0.36)\,mb \,
\label{eq:16}
\endarr
and a small multiplicity of neutrons
$\langle n_{n}^{(2)} \rangle = {2\over 9}$.
Inspection of Eq.(\ref{eq:11}) shows that
$\sigma_{SD}(N\pi+N\pi\pi) \approx 0.7-0.8\,mb$
practically exhausts the contributions from
$M_{Y} \lsim 2.3\,GeV$.
For the excitation of the continuum of still
higher masses $5 \lsim M_{Y}^{2} \lsim 25 GeV^{2}$,
Eq.~(\ref{eq:11}) gives an
estimate of $\sigma_{3} \approx 1.1-1.4\,mb$.
There are no direct experimental data on the spectrum of
neutrons in the continuum HM excitation. However, because of a well
known similarity of secondary particle spectra
in the pomeron-proton and hadron-proton collision at
a similar center-of-mass energy \cite{K79,AG81,G83,ZZ88}, production
of  neutrons from HM excitations can be approximated by
a meson-exchange model \cite{ZS84,HSS,HLNSS94},
which gives a good description of the hadronic production data.
In the specific model \cite{HSS,HLNSS94} we find a mean multiplicity
of neutrons $\langle n_{n}^{(3)} \rangle = 0.2$.

Putting together the three components discussed above,
we find an average multiplicity of neutrons in double diffraction
dissociation
\begin{equation}
\langle n_{n}^{DD} \rangle \equiv
{\sum_i \langle n_{n}^{(i)} \rangle \sigma_{i} \over \sum \sigma_{i} }
\approx 0.30-0.31 \; .
\end{equation}
Thus the expected fraction of neutron tagged large rapidity gap events
equals
$f_{DD}^{(n)} =
\langle n_{n}^{DD} \rangle \cdot f_{DD} \approx 0.08$.

Experimental cuts on the neutron energy make the evaluation
of the $z$-distribution of the produced neutrons very important.
Let us start the evaluation with the $n\pi$ system assuming full
angular coverage for the outgoing neutrons.
Let $\theta_{GJ}$ be the so-called
Gottfried-Jackson angle between the neutron and the $n\pi$
motion axis. Then
\beq
\Phi_{1}(z) = \int d\cos\theta_{GJ}dM_{Y}\,
{1 \over N_{ev}}\cdot {dN_{ev} \over dM_{Y}d\cos(\theta_{GJ})}
\delta(z-
{E^{*}+p^{*}\cos\theta_{GJ}\over M_{Y}})\,.
\label{eq:18}
\endeq
(The normalization is such that $\int_{0}^{1} dz \Phi_{1}(z)=1$).
Notice the useful lower bound  $z \gsim m_{n}^{2}/M_{Y}^{2}$,
which shows that all the neutrons from excitations with
$M_{Y}^{2} \lsim  m_{n}^{2}/z_{min}$
pass the $z_{min}$ cut $z \gsim z_{min}$.
With $z_{\min}\approx 0.5$ \cite{FNC} one has to worry only about
the loss of neutrons from excitations with
$M_{Y}^{2} \gsim 2 GeV^{2}$. At such $M_{Y}$, the
$\cos\theta_{GJ}$ distribution measured \cite{B78} exhibits a
strong forward peak at $\cos\theta_{GJ} = 1$, which results in a peak
at $z\sim 1$, and a less prominent backward peak at
$\cos\theta_{GJ} = -1$, which contributes to
$z \sim m_{n}^{2}/M_{Y}^{2}$.
The results of the evaluation of $\Phi_{1}(z)$ using the experimental
data on the $(M_{Y},\cos\theta_{GJ})$ distribution of $N_{ev}$
from Ref. \cite{B78} are presented in Fig.~2a (solid line).
The $\Phi_{1}(z)$ distribution peaks at $z \sim 1$.
A comparison with a flat $\cos\theta_{GJ}$ distribution (dashed line)
shows a rather weak dependence of the loss of neutrons on
a detailed form of the $\cos\theta_{GJ}$ distribution.
We find that a $\epsilon_{n}^{(1)}\approx 0.85$
fraction of neutrons from $n \pi^+$ excitations of the proton
passes the $z_{min}=0.5$ cut.

One can repeat a similar anlysis for the $n\pi\pi$ excitation,
modelling it by $\Delta\pi$ production (see \cite{W75,C80}).
Again we can use the experimentally measured mass distribution
as given for instance in Ref.\cite{AG81}.
The $z$-spectra of neutrons are fairly independent of the angular
distributions in the Gottfried-Jackson frame. We assume the
$cos \theta_{GJ}$ distribution
for the $\Delta \pi$ system the same as for
the $N \pi$ component and take isotropic distribution
in the $\Delta \rightarrow n+\pi$ decay.
While the $z$ distribution of primary $\Delta's$ is very similar
to $\Phi_{1}(z)$, the $z$ distribution of neutrons
after the $\Delta \rightarrow n\pi$ decay
becomes peaked at $z \sim 0.6$ (Fig.~2b).
Even so, the loss of neutrons from $n \pi \pi$ excitations of
the proton is not strong and
we find $\epsilon_{n}^{(2)}\approx 0.71$
for the $z_{min}=0.5$ cut.

For the $z$-distribution of fast neutrons from HM excitations
we take the spectrum for non-diffractive DIS calculated and tested
against hadronic experimental data in Ref.\cite{HLNSS94}.
At $z \gsim 0.5$ it is dominated by the familiar pion exchange
mechanism.
The results for $\Phi_{3}(z)$ are presented in  Fig.~2b. We
find $\epsilon_{n}^{(3)}\approx 0.64$
for the $z_{min}=0.5$ cut.
Finally, putting together all the three components,
we find
$\epsilon_{n}^{DD}(z_{min}=0.5) \equiv
{\sum_i \epsilon_{n}^{(i)}(z_{min}=0.5)
\langle n_{n}^{(i)} \rangle \sigma_{i}
\over
\sum_i \langle n_{n}^{(i)} \rangle \sigma_{i} }
\approx 0.75$.

It is interesting to compare the neutron production
in diffractive and non-diffractive DIS. In the latter case
the major source of forward neutrons is the pion exchange with deep
inelastic scattering on pions (for a detailed formalism see
Ref.\cite{HSS} and for a discussion \cite{HLNSS94}).
For small $x$ relevant at HERA, we can use the pion structure
function from GRV model \cite{GRV92},
which gives
$F_{2}^{\pi}(x,Q^{2})
\approx$ (0.7-0.8)$ F_{2}^{p}(x,Q^{2})$.
The resulting spectrum of neutrons from non-diffractive DIS
$\Phi^{DIS}(z)$ (neutrons from $\Delta$ decays included) is
shown in Fig.~3a.
The average multiplicity of neutrons
in non-diffractive DIS at small $x$ equals
$\langle n_{n}^{DIS}\rangle =  0.15 $ and
a $\epsilon_{n}^{DIS}= 0.64$ fraction of neutrons passes the
$z_{min} = 0.5$ cut.

Notice, that the spectrum of neutrons from DD events is peaked
at larger $z$ in comparison to non-diffractive DIS,
which suggests an interesting possibility
to enhance the observed fraction of large
rapidity gap events by increasing the $z_{\min}$ cut, at the
expense of losing a part of statistics.
The $z$-cut dependence of
$\epsilon_{n}^{DD}(z_{min})$ and $\epsilon_{n}^{DIS}(z_{min})$
for DD and non-diffractive DIS, respectively, is shown in Fig.~3b, and
the gain factor
$
R_{n}(z_{min}) \equiv
\epsilon_{n}^{DD}(z_{min})/\epsilon_{n}^{DIS}(z_{min})
$
is shown in Fig.~3c.

Up to now we have considered the ideal case of full angular acceptance
for outgoing neutrons. In realistic case of FNC the angular acceptance
in the polar angle $\theta$ is finite ( $\theta_{m}$ ).
This means that only neutrons with transverse momentum
$k_{\perp} \leq zk_{\perp,m}=zE_{p}\theta_{m}$ can reach the FNC.
The $k_{\perp}$-acceptance rises with $z$.
For a Gaussian $k_{\perp}$-distribution, the effect of the angular
cut can be included by the simple substitution
\beq
\Phi_{i}(z) \longrightarrow \Phi_{i}(z)
\left[1-\exp\left(-{z^{2}k_{\perp,m}^{2} \over
\langle k_{\perp}^{2}(z)\rangle } \right)
\right]\,.
\label{eq:21}
\endeq
The mean transverse momentum squared of neutrons
$\langle k_{\perp}^{2}(z)\rangle $ can, in principle,
be directly measured by FNC. The experimental information
on $\langle k_{\perp}^{2}(z)\rangle $ is contained in the
data from the hadronic experiments, although it is not cited directly
in the relevant experimental publications. For a rough estimate,
we take $\langle k_{\perp}^{2}(z)\rangle = 0.25\,GeV^{2}$ and
$\theta_{m}=7\cdot 10^{-4}$, which is
relevant for the test FNC of the ZEUS collaboration \cite{FNC}.
The effect of such an angular cut on the neutron spectra is
is shown in Figs.~4.
It somewhat enhances the observed DD signal compared to the
signal of non-diffractive DIS.

Neutron tagging is an efficient tool for testing
the mechanism  of DD. For instance, zooming at neutrons with
$z\sim 1$ suppresses the "inelastic" background in the
$J/\Psi$ production and allows a more accurate determination
of $B_{el}(J/\Psi)$. On the other hand, zooming at
$z\sim 0.5$ would enhance the inelastic background and
enable a determination of $B_{in}(J/\Psi)$, which is also a
quantity of theoretical interest.

\vskip 1cm

\section{ Conclusions}

Based on the Regge factorization, we
find a substantial contribution of double
diffraction dissociation to exclusive production of
vector mesons and large rapidity gap events at HERA.
In diffractive excitations of virtual photons into low mass states
the DD background fraction is
$f_{DD}(\gamma^* \rightarrow X, M_{X}^2 \sim Q^2) \approx 0.21-0.23$,
whereas for high mass excitations of the photon
the double diffraction dissociation background is somewhat larger
$f_{DD}(\gamma^* \rightarrow X, M_{X}^2 \gg Q^2) \approx 0.26-0.28$.
Therefore carefull corrections for DD background are neccessary for
the interpretation of the large rapidity gap data in terms of
the $\beta$-dependence of the pomeron structure function.

In real photoproduction of light vector mesons we expect an inelastic
background $f_{in}(\gamma \rightarrow \rho^0) \sim 0.21-0.23$.
In contrast, in the $J/\Psi$ photoproduction
the inelastic background is expected to be stronger:
$f_{in}(J/\Psi) \sim 0.26-0.28$.

The presented simple estimates confirm the
feasability of tagging the double diffraction events by fast
neutrons from products of diffraction excitation of the proton.
Including effects of finite angular acceptance, we have estimated
the fraction of double diffraction dissociation events in the large
rapidity gap sample to be $f_{DD}\approx 0.21-0.23$. Approximately
25\% of the double diffraction dissociation events contain
a neutron with $E_{n}/E_{beam} > 0.5$.

We find that the spectrum of neutrons from DD events
peaks at $z$ larger than that for non-diffractive DIS,
which should facilitate their experimental
observation. The experimental determination of the $Q^{2}$
and $x_{\Pom}$ dependence of the spectrum and multiplicity
of neutrons in the large rapidity gap events
(including real photoproduction), and comparison
with the related proton-proton interactions,
will be of great importance for testing the factorization properties
of the pomeron in an entirely new kinematical domain of
deep inelastic scattering.
\medskip\\

{\bf Acknowledgements:} One of us (A.S.) thanks Garry Levman
for very interesting discussion which stimulated this work.
The work was performed during the stay of A.S.
at the Institut f\"ur Kernphysik, KFA, J\"ulich.
This work was partly supported by the INTAS Grant 93-239.

\pagebreak

\pagebreak

{\bf Figure captions:}\medskip\\

\begin{itemize}

\item[Fig.~1]
- - - Single and double diffraction dissociation in DIS and
proton-proton scattering.

\item[Fig.~2]
- - - a) The normalized spectrum  of neutrons for the $N\pi$ component
for the experimentally observed forward-backward peaked
(solid curve) and flat (dashed curve) $\cos\theta_{GJ}$
distributions.\\
- - - b) Normalized spectra of neutrons $\Phi_{i}(z)$ for the $N\pi$
(solid curve) , $N\pi\pi$ (dashed curve) and the high mass (dotted
curve) components of DD.

\item[Fig.~3]
- - - a) Spectrum of neutrons from diffractive (solid) and non-diffractive
     (dashed) DIS normalized to unity. \\
- - - b) The fraction of neutrons  with $z\gsim z_{min}$
    (tagging efficiency) for DD
$\epsilon_{n}^{DD}(z_{min})$ (solid curve) and non-diffractive DIS
$\epsilon_{n}^{DIS}(z_{min})$ (dashed curve). \\
- - - c) The gain factor $R_{n}(z_{min})$ with full angular acceptance. \\

\item[Fig~4]
- - - The effect of the finite angular acceptance of FNC on the
neutron spectra from diffractive (solid) and non-diffractive
(dashed) DIS.

 \end{itemize}
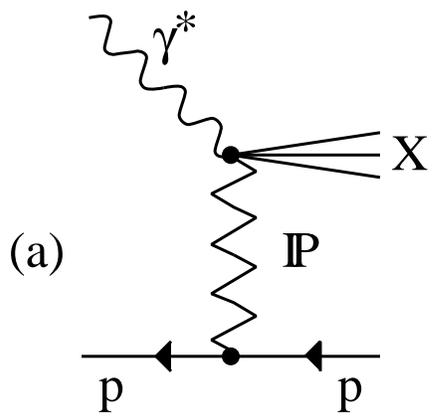
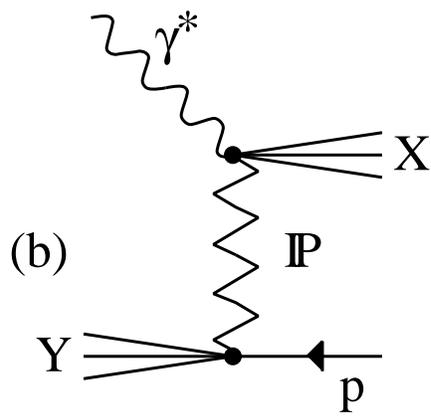
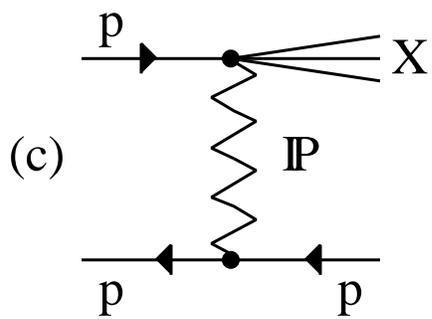
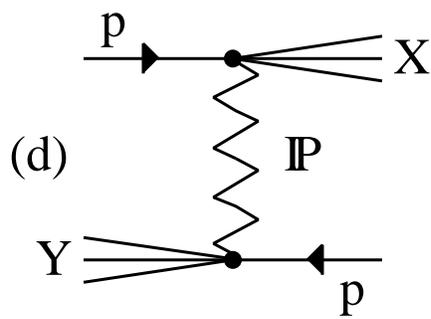
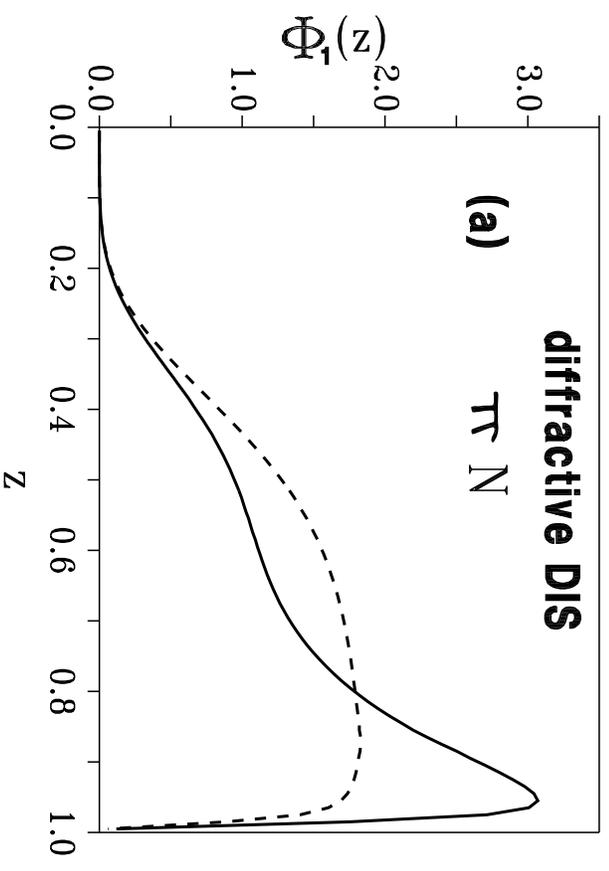
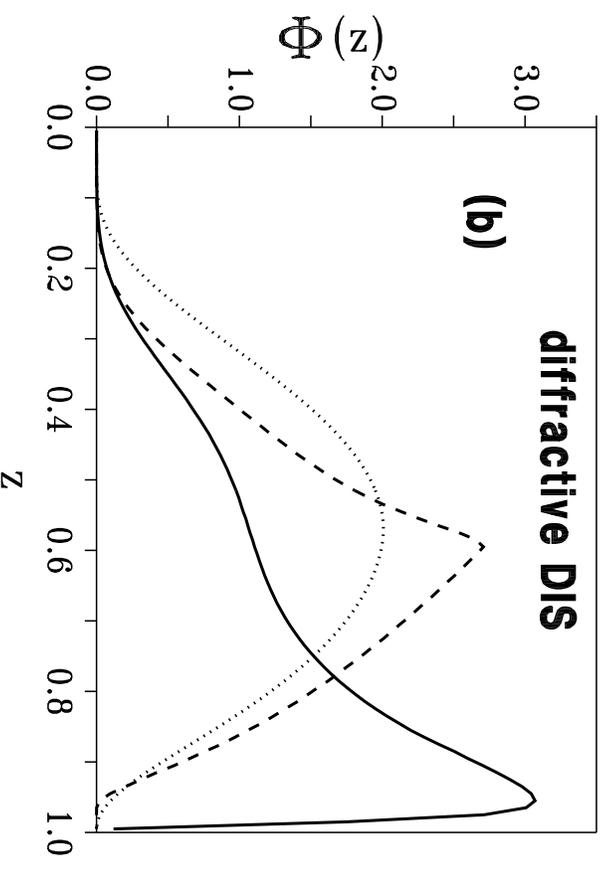
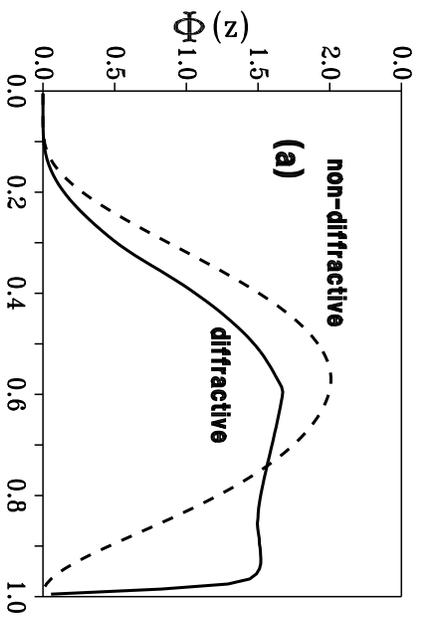
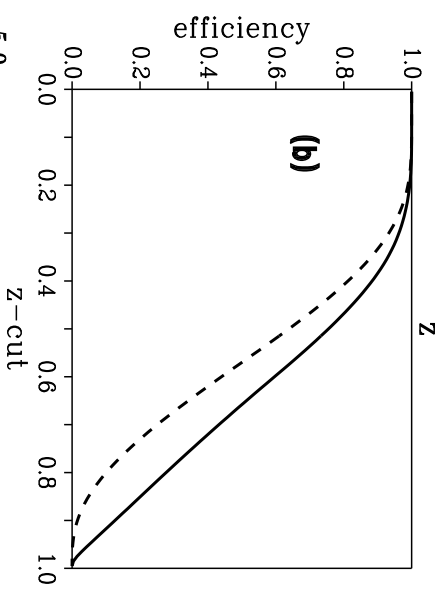
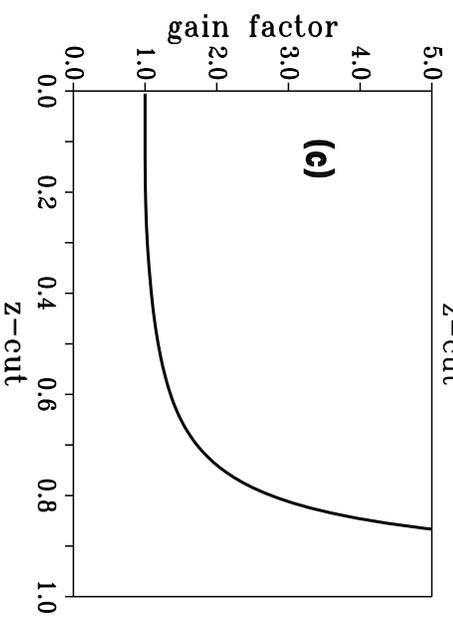
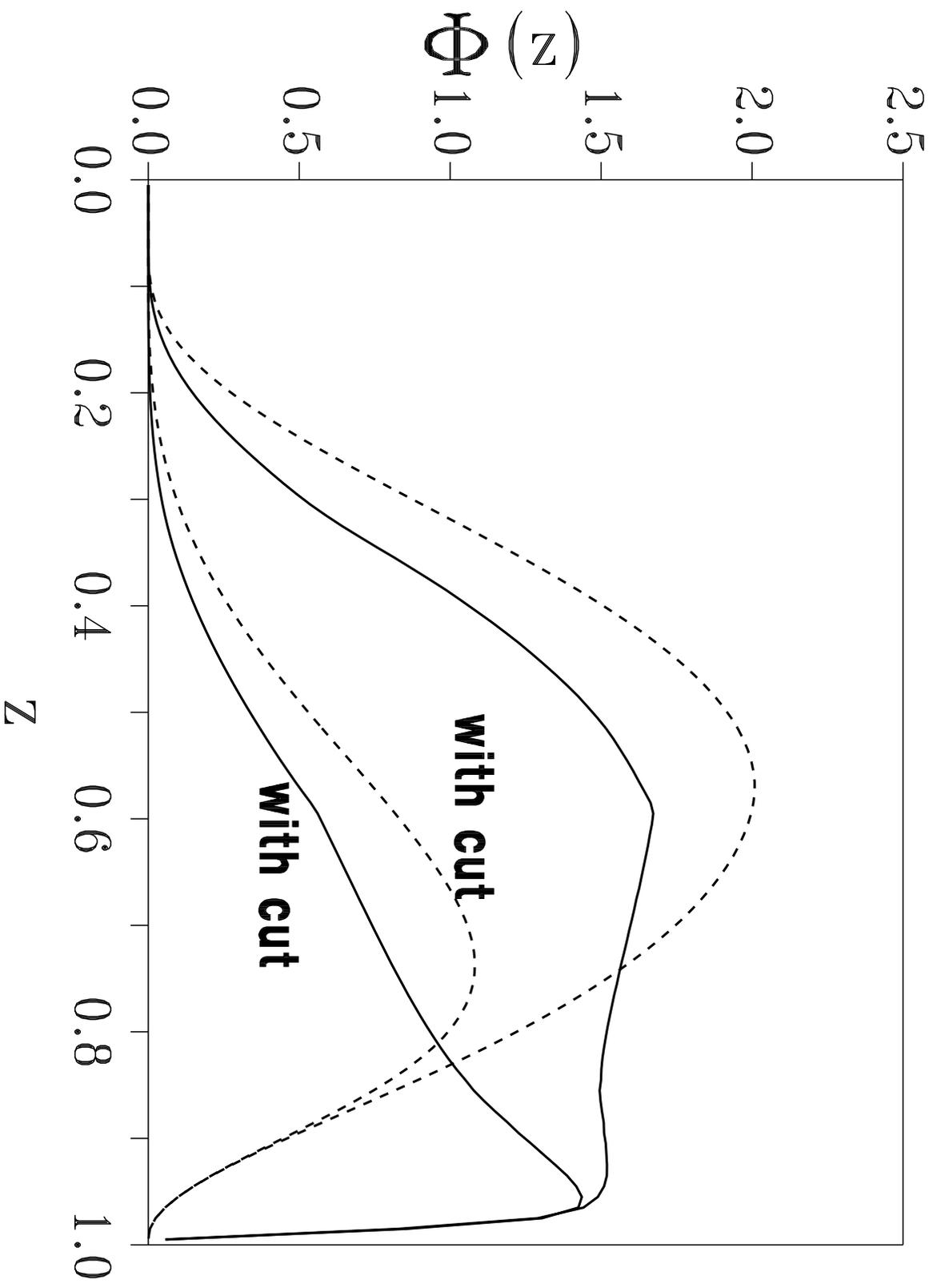
\end{document}